\documentclass[aps,prr,preprint,groupedaddress,amsmath,epsfig,amssymb,eqsecnum]{revtex4} 
\usepackage{makecell}
\usepackage{bbm}
\usepackage{mathrsfs}
\usepackage{subfigure}
\usepackage{enumerate}
\usepackage{amsfonts}
\usepackage{color}
\usepackage{parskip}
\usepackage{graphicx}
\newcommand{\dbar}{d\hspace*{-0.08em}\bar{}\hspace*{0.1em}}
\usepackage{relsize}

\newcommand{\be}{\begin{equation}}
\newcommand{\ee}{\end{equation}}
\newcommand{\ba}{\begin{eqnarray}}
\newcommand{\ea}{\end{eqnarray}}

\makeatletter
\newdimen\scalemath@axis
\newcommand*{\scalemath}[3]{%
  #1{%
    \mathpalette{\scalemath@aux{#2}}{#3}%
  }%
}
\newcommand*{\scalemath@aux}[3]{%
  \begingroup
    \everyvbox{}%
    \settoheight\scalemath@axis{$#2\vcenter{}$}%
    \raisebox{\scalemath@axis}{%
      \scalebox{#1}{%
        \raisebox{-\scalemath@axis}{%
          $\m@th#2#3$%
        }%
      }%
    }%
  \endgroup
}
\makeatother 

\begin{document}
\title{A Rigorous Foundation for Stochastic Thermodynamics \\
via the Microcanonical Ensemble}
\author{Xiangjun Xing$^{1,2,3}$}
\email{xxing@sjtu.edu.cn}
\address{$^1$Wilczek Quantum Center, School of Physics and Astronomy, Shanghai Jiao Tong University, Shanghai 200240, China \\
$^2$T.D. Lee Institute, Shanghai Jiao Tong University, Shanghai 200240, China\\
$^3$Shanghai Research Center for Quantum Sciences, Shanghai 201315, China}
\date{\today} 

\begin{abstract}
We consider a small Hamiltonian system strongly interacting with a much larger Hamiltonian system (the bath), while being driven by both a time-dependent control parameter and non-conservative forces. The joint system is assumed to be thermally isolated. Under the assumption of time-scale separation (TSS)---where the bath equilibrates much faster than the system and the external driving---the bath remains in \emph{instantaneous equilibrium}, described by the \emph{microcanonical ensemble} conditioned on the system state and the control parameter.  We identify a decomposition of the total Hamiltonian that renders the bath energy an adiabatic invariant under slow evolution. This same decomposition defines the system Hamiltonian as the \emph{Hamiltonian of mean force}, and ensures that neither the system nor the control parameter does reactive work on the bath. Using time-reversal symmetry and TSS, and without invoking any model details, we rigorously prove that the reduced dynamics of the system is Markovian and satisfies a form of \emph{local detailed balance} (LDB) which involves transition probabilities but not path probabilities.  By working entirely within \emph{ the microcanonical framework} and adopting a precise decomposition of the total energy, we provide rigorous definitions of bath entropy as the Boltzmann entropy, and of heat as  the negative change of the bath energy.  Our approach bypasses the ambiguities associated with conventional definitions of thermodynamic variables and path probabilities, and establishes a rigorous and thermodynamically consistent foundation for stochastic thermodynamics, valid even under strong system–bath coupling.
\end{abstract}

\maketitle 

\section{Introduction}
\label{sec:intro}

Stochastic thermodynamics~\cite{Peliti-book,sekimoto1998,Jarzynski-review,Seifert-review} provides a powerful framework for non-equilibrium physics of small systems coupled to fluctuating environments. Two cornerstones  of this framework are the \emph{Markovianity}  and the \emph{local detailed balance} (LDB). Markovianity allows us  to ignore the detailed dynamics of the surrounding bath and to assign consistent probabilities to system trajectories.   The LDB principle, in turn, provides the essential link between Markovian dynamics and the thermodynamics of the heat bath. Together, these two properties underpin entropy production~\cite{seifert2005}, fluctuation theorems~\cite{gallavotti1995,lebowitz1999,Crooks1999,Jarzynski1997,Bochkov1977,Bochkov1981}, and other major results in stochastic thermodynamics.

The Markovian property of reduced dynamics is usually justified using projection operator techniques such as the Zwanzig--Mori formalism~\cite{zwanzig1960,mori1965,Zwanzig-book}, where weak coupling and near-equilibrium assumptions are introduced to facilitate explicit computations. However, in recent years, it has been gradually realized that these assumptions are not necessary to guarantee the emergence of Markovian dynamics~\cite{Breuer-2005, Hanggi-2005,Xing-CG}.  The essential requirement is \emph{time-scale separation} (TSS), which ensures that the bath remains in \emph{ instantaneous equilibrium} conditioned on the system’s state and control parameters.  Despite this growing awareness, a model-independent derivation of  Markovianity from TSS has remained lacking.  

The principle of LDB is usually presented as an extension of the equilibrium detailed balance (DB) condition for the transition probabilities~\cite{bergmann1955}:
\begin{equation}
P(x'|x; t)\, p_{\text{eq}}(x) = P(x|x'; t)\, p_{\text{eq}}(x'),
\label{DB-1}
\end{equation}
which, under weak coupling and bath equilibrium, can be recast into a form involving exchanged heat or change of bath entropy. Further assuming Markovian property (which ensures instantaneous bath equilibrium), this can be adapted to the LDB condition for non-stationary processes with time-dependent control parameter~\cite{Crooks1999}:
\begin{equation}
\frac{P_F[\gamma|\gamma_0]}{P_B[\tilde\gamma|\tilde\gamma_0]} = e^{-\beta Q[\gamma]} = e^{\Delta S_Y[\gamma]},
\label{LDB-gamma}
\end{equation}
where \(P_F[\gamma|\gamma_0]\) and \(P_B[\tilde\gamma|\tilde\gamma_0]\) are the conditional probabilities of a forward trajectory and its time-reversed counterpart, and \(Q[\gamma]\) is the heat transferred to the bath, whereas  \( \Delta S_Y[\gamma] \) is the entropy change of the bath~\footnote{Some author prefer an ``unconditional'' version of the LDB condition, in which both the forward and backward processes are assigned fixed initial distributions. While this approach is convenient for deriving certain fluctuation theorems, it introduces additional complications regarding the choice of initial conditions and the interpretation of trajectory ensembles, and somewhat obscures the essential new physics contained in LDB, which is the link between the system dynamics and the bath thermodynamics.  We shall therefore focus on the conditional version of LDB in the present work.}.    Note that we take the convention $k_B = 1$, so that $\beta = 1/T$.     While Eq.~(\ref{LDB-gamma}) has been widely adopted, it presumes weak interaction and Markovianity, that $Q[\gamma]$ and $S_Y[\gamma]$  are inferred from system dynamics, and that the path probabilities can be very subtle to define rigorously.  Under strong coupling~\cite{Seifert-SC,Hanggi-2016-SC,Jarzynski-2017-SC,Esposito-2017-SC,Hanggi-2020-comment,Hanggi-2020-Colloquium,Xing-SC-1,Xing-SC-2} or in the presence of multiplicative noises~\cite{{Xing-PI}}, Eq.~(\ref{LDB-gamma}) may lose precise meaning and can become dangerously misleading.  
   
A more formal derivation of LDB was supplied by Maes and Netocny~\cite{Maes2003,Maes2020}, who relate the ratio of coarse-grained transition probabilities arising from an underlying time-reversible unitary dynamics to the difference in Boltzmann entropy (of the total system) between the final and initial states. A careful inspection indicates that these authors assumed the stationary nature of the dynamics.   More importantly, their general formulation does not clarify the connection between Boltzmann entropy change and heat exchange between the system and its reservoirs, and thus remains detached from the concrete thermodynamic structure of physical systems.   As Maes himself acknowledged in his later work~\cite{Maes2020}, additional assumptions such as weak coupling and near-equilibrium conditions are required to apply the formalism to realistic systems. Hence the approach of Refs.~\cite{Maes2003,Maes2020} shares the same limitations as the path-probability-based derivation of LDB due to Crooks~\cite{Crooks1999}. 

At the heart of many ambiguities in stochastic thermodynamics lies the fact that the thermodynamics of the bath is not explicitly formulated---a direct consequence of the popular canonical ensemble approach. As a result, heat, bath entropy change, and entropy production must be inferred indirectly from the system dynamics, often leading to ambiguity or possible inconsistency.  This ambiguity is substantially amplified if the coupling between the system and the bath is not fixed but needs to be tuned.  For example, in Jarzynski’s original derivation of the nonequilibrium work relation,  the system is assumed to be initially in thermal equilibrium with a heat bath, but the bath is then disconnected during the subsequent evolution in order to invoke Liouville’s theorem. This procedure is potentially dangerous: there is always legitimate doubt about subtle or uncontrolled effects that may arise when a bath is coupled or decoupled.  While Jarzynski’s maneuver is mathematical in nature and may be circumvented by alternative methods of proof, there are physical processes---such as cyclic thermodynamic engines---where the tuning of the system-bath coupling is real and plays an essential role. In such cases, the ambiguities introduced by the canonical ensemble approach cannot be avoided.

It is therefore conceptually advantageous to adopt the microcanonical ensemble, where the total system---including both the bath and the system---is modeled explicitly as an isolated Hamiltonian system, the bath thermodynamics is derived rather than assumed, and the system-bath coupling is treated explicitly.  It is this more difficult and rarely pursued path that we shall follow in the present work, with the aim of establishing both Markovianity and LDB within a single, unified framework, irrespective of interaction strength and other model details. 

We note in passing that the microcanonical ensemble has been extensively studied in the context of fluctuation theorems and thermalization of small isolated quantum systems~\cite{Talkner2008,Talkner2013,Polkovnikov2011,Goold2016,Vinjanampathy2016,Gogolin2016}. Related questions have also been raised concerning the validity of thermodynamic laws in such isolated small systems~\cite{Hanggi-2014-entropy,Hanggi-2016-entropy,Frenkel-2015-entropy}. In contrast, our use of the microcanonical ensemble does not stem from an assumption of isolation, but from the need to treat \emph{thermal interactions between system and bath explicitly and consistently}. We begin from the empirically justified premise that thermodynamic laws hold for large isolated composite systems---comprising a small subsystem coupled to a large environment---in the thermodynamic limit, up to sub-extensive corrections. The thermodynamics of small systems, in this view, does not emerge from their behavior in isolation, but from a careful analysis of the composite dynamics, followed by a principled reduction to the subsystem via coarse-graining and time-scale separation. This perspective retains both conceptual clarity and physical realism, especially in regimes of strong coupling and time-dependent driving, where traditional approaches—based on either the microcanonical or canonical ensemble—often break down.

The logic behind our approach also closely parallels Leggett's treatment~\cite{Leggett-book} of superfluidity and superconductivity using the canonical ensemble. Just as Leggett insisted on conserving particle number to reveal the true nature of condensation, we insist on modeling the total system with a microcanonical ensemble to preserve energy conservation and clarify the physical origins of stochastic irreversibility and entropy production. This approach, though technically more involved, is far more physically illuminating than the conventional canonical ensemble approach. 

 We make the following assumptions: (i) The bath dynamics is much faster than the system dynamics (TSS); (ii) the size of the bath is much larger than that of the system; and (iii) the joint dynamics is time-reversal symmetric, and the coarse-grained joint dynamics is Markovian.   We model the combined system---comprising the system of interest ($X$) and its environment ($Y$)---as an isolated Hamiltonian system, with a clear separation of the total Hamiltonian into a system part and a bath part.  As a consequence, both the system energy and the bath energy are well defined, and the heat $Q$ is defined as the energy transfer from the bath to the system.   Because of TSS, the bath remains in instantaneous equilibrium, which is described by the \emph{microcanonical ensemble}, conditioned on the system state and  control parameter.   The entropy of the bath $S_Y$ is defined as its Boltzmann entropy. These definitions remain valid regardless of the strength of interaction and correlation between the system and the bath.  We then rigorously show that the system dynamics is Markovian and its transition probabilities satisfy a LDB condition:
\begin{equation}
\frac{P_F(Q, x_t | x_0; \beta)}{P_B(-Q, x_0^* | x_t^*; \beta)} 
= e^{-\beta Q} = e^{\Delta S_Y},
\label{LDB-1-0}
\end{equation}
where $x_0^*,x_t^*$ are respectively the time reversal of $x_0,x_t$. 

{One important clarification is in order regarding our treatment of Markovianity. What we establish in this work is that, under time-scale separation (TSS), the \emph{Markovian property of the reduced $X$ dynamics follows from that of the coarse-grained joint $(X,Y)$ dynamics}. The Markovianity of the joint dynamics is \emph{assumed}, not derived from microscopic principles. While this assumption is physically well-motivated by the rapid mixing of the bath, a general derivation of Markovianity under coarse-graining of deterministic Hamiltonian systems remains an outstanding and difficult open problem. Our result, therefore, should not be understood as a complete derivation of Markovian stochasticity from first principles. }

The significance of our results lies in four key advances. First, by formulating the theory within the microcanonical ensemble, we avoid the ambiguities inherent in conventional treatments of bath thermodynamics, where heat and entropy must be inferred indirectly from the system. Second, our derivation avoids the problematic use of path probabilities and instead relies solely on well-defined transition probabilities. Third, all results follow rigorously from two physical principles---time-reversal symmetry and time-scale separation---without requiring any specific model details. Finally, our framework fully extends the scope of stochastic thermodynamics to strongly coupled systems and tunable system-bath interactions, provided the total Hamiltonian is decomposed wisely and the bath entropy is interpreted thermodynamically via Boltzmann. Together, these features not only consolidate the theoretical foundation of stochastic thermodynamics, but also substantially broaden its domain of applicability.

To put our results into a concrete setting, we consider a well-konwn example of a colloid X immersed in an equilibrium fluid Y of macroscopic size. The interaction between the colloid and the fluid is strong but is limited within a thin interfacial layer. The parameter $\lambda$ may control the  properties of the interface between the colloid and the fluid, but does not influence the bulk properties of the fluid.   As we move the colloid, or tune the control parameter, the fluid remains in an instantaneous equilibrium, with negligible changes in intensive parameters, such as the temperature, or energy density, pressure, etc, which characterize the bulk properties of the fluid.   Conversely, the bulk properties of the fluid influences the static and dynamic properties of the colloid only indirectly through the these aforementioned intensive parameters.   For example, if we add an energy $\Delta E$ to the fluid, which is large compared to the energy of the colloid but small comparing with the energy of the fluid, the resulting change of the static or dynamic properties of colloid is negligible.  These intuitively obvious results will play an important role in our discussion below.  In this particular case, the Markovianity and the LDB (\ref{LDB-1-0}) means that the reduced dynamics of the colloid can be described by a nonlinear Langevin equation with white noises and satisfying DB conditons, whose most general form of such equation is described in detail in  Ref.~\cite{Xing-NC-ST}. 

The remainder of this paper is organized as follows.  In Section~\ref{sec:decomp-H} we construct a decomposition of the total Hamiltonian such that neither the system state variable nor the control parameter does reactive work on the bath.  This ensures that the bath energy is an adiabatic invariant, and that the system Hamiltonian is the Hamiltonian of mean force. In Section~\ref{sec:Markov}, we use TSS and the extensive nature of the bath to prove the Chapman--Kolmogorov equation for the reduced dynamics, thereby establishing the Markov property. In Section~\ref{sec:W-Q} we define heat and work rigorously and confirms their consistency with previous results. In Section~\ref{sec:LDB}, we use conditional equilibrium and adiabatic invariance to derive the LDB relation. In Section~\ref{sec:FTs} we demonstrate the power of our formulation by deriving a family of fluctuation theorems using the microcanonical ensemble theory.  To the best of our knowledge, this is the first systematic proof of fluctuation theorems using the microcanonical ensemble.   Furthermore, some of these FT that involve initial and/or final system states turn out to be novel.  Finally, in Section~\ref{sec:conclusion} we connect our results with some of our previous works and hilight the significance of our approach from the viewpoint of statistical ensemble theory.

\section{Adiabatic Invariance of Bath Hamiltonian}
\label{sec:decomp-H}

Consider an isolated Hamiltonian system consisting of slow variables \( X \)  (the small system) with a generic state \( x = (q_x,p_x)\) and fast variables \( Y \) (the large bath) with a generic  state \( y = (q_y, p_y) \).   We decompose the total Hamiltonian in the following way:
\ba
H_{\text{tot}}(x, y, \lambda) = H_X(x; \lambda) + H_Y(y; x,\lambda),
\label{H_tot-decomp}
\ea
where $H_X$ and $H_Y$ are respectively  the system Hamiltonian and the bath Hamiltonian.  Note that $\lambda$ appears both in $H_X$ and in $H_Y$.  As we have emphasized in the Introduction via the example of a colloid in a fluid, $\lambda$ does not couple to the bulk properties of $Y$.  Except the fact the system Hamiltonian $H_X(x; \lambda)$ is independent of the bath variable $y$, Eq.~(\ref{H_tot-decomp}) does not uniquely determines the decomposition of the total Hamiltonian.  We shall choose the particular decomposition that makes the reduced thermodynamic theory of the system as simple as possible.

According to TSS, Y is in a microcanonical equilibrium conditioned on $x$ and $\lambda$, which is described by an equal probability distribution on the energy surface:
\ba
p_Y(y; E_Y, x,\lambda) = \frac{ \delta(H_Y(y; x,\lambda) - E_Y )}
{\Omega_Y(E_Y; x,\lambda) },  
\label{p_Y-cond-EQ}
\ea
 where
 \ba
\Omega_Y(E_Y; x,\lambda) \equiv 
\int dy\, \delta (E_Y - H_Y(y; x,\lambda)),
\label{Omega-def}
\ea
 is the microcanonical partition function, which is related to the Boltzmann entropy of $Y$ via 
\ba
S_Y(E_Y; x,\lambda) = \log \Omega_Y(E_Y; x,\lambda). 
\label{S_Y}
\ea 
The differential form of $S_Y(E_Y; x,\lambda)$ is
\ba
d S _Y
= \beta dE_Y  
+  \left[\frac{\partial S _Y}{\partial x}\right]_{E_Y,\lambda} \!\!\!\! dx 
+ \left[\frac{\partial S _Y}{\partial \lambda}\right]_{E_Y, x} \!\!\!\! d\lambda , \quad
\label{dS_Y-neq-0}
\ea
where $\beta = 1/T = \left[{\partial S _Y}/{\partial E_Y}\right]_{x,\lambda} $ is the inverse temperature.  We further rewrite Eq.~(\ref{dS_Y-neq-0}) into 
\ba
dE_Y   = T\, dS_Y -T  \left[\frac{\partial S _Y}{\partial x}\right]_{E_Y,\lambda} \!\!\!\! dx  
-T  \left[\frac{\partial S _Y}{\partial \lambda}\right]_{E_Y, x} \!\!\!\! d\lambda ,
\quad \label{dE_Y-neq-0}
\ea
which can be understood as the first law applied to the bath.  The first term in the r.h.s. must be deemed as \emph{ negative the heat transfer from the bath to the system}, whereas the remaining two terms are the works done to the bath,  by the system and by the external control parameter respectively. 

The possibility that the control parameter \( \lambda \) performs work on the bath does not contradict any thermodynamic law. However, it obstructs the construction of a reduced thermodynamic theory that depends only on the system variables.  More subtly, if changes in the system state also induce work on the bath, then the bath must exert a reactive force on the system. In Langevin-type descriptions---such as the dynamics of a colloidal particle---this reactive force becomes intertwined with dissipative forces originating from the bath fluctuations, making it difficult to disentangle reversible and irreversible contributions.  Hence, the thermodynamic theory is much simpler if neither the control parameter nor the system does any work to the bath.

Our goal is therefore to find the particular decomposition of Hamiltonian (\ref{H_tot-decomp}) such that 
\ba
\frac{\partial S _Y}{\partial \lambda}  = \frac{\partial S _Y}{\partial x} = 0. 
\label{dE_Y-eq-0}
\ea
 An equivalent representation of Eqs.~(\ref{dE_Y-eq-0}) is 
\ba
  \frac{\partial \Omega_Y}{\partial x}  = \frac{\partial \Omega_Y}{\partial \lambda} = 0. 
\label{Omega_Y-eq-0-2}
\ea 
As a consequence, the bath energy depends only on $S_Y$ but not on $x, \lambda$, which means that it is \emph{ an adiabatic invariant}---that is, it remains invariant under slow variations of the system state and the control parameter.   The corresponding bath Hamiltonian is then called \emph{ an adiabatically invariant Hamiltonian}.  We can then rewrite Eqs.~(\ref{p_Y-cond-EQ}) and (\ref{S_Y}) as
\begin{subequations}
\label{Omega-S_Y-indep}
\ba
\Omega_Y(E_Y)  &=& \log \int dy\, \delta (E_Y - H_Y(y; x,\lambda)),
\\
S_Y(E_Y )  &=& \log\Omega_Y(E_Y) ,  
\ea
\end{subequations}
and rewrite Eq.~(\ref{dE_Y-neq-0}) as
\ba
d E_Y = T\, dS_Y = - \dbar Q,  \label{dE-S-Q-Y}
\ea
which means that the heat $\dbar Q$ is just negative the change of the bath energy.   Equation (\ref{dE-S-Q-Y}) will play a significant role in Sec.~\ref{sec:LDB}, where we prove rigorously LDB.

\subsection{Construction of Bath Hamiltonian}
Let us now show explicity how to construct an adiabatically invariant Hamiltonian for the bath.  Suppose we have an arbitrary decomposition:
\ba
H_{\text{tot}}(x, y, \lambda) = H^0_X(x; \lambda) + H^0_Y(y; x,\lambda),
\label{H_tot-decomp-0}
\ea
The microcanonical partition function and Boltzmann entropy of $Y$, with respect to $H_Y^0(y; x,\lambda)$, are given by
\ba
\Omega^0_Y(E_Y; x,\lambda) &=& 
\int dy\, \delta (E_Y - H_Y^0(y; x,\lambda)),
\label{Omega-0-def}\\
S^0_Y(E_Y; x,\lambda) &=& \log \Omega^0_Y(E_Y; x,\lambda). 
\label{S^0_Y}
\ea
both of which generally depend on $x,\lambda$.    We  choose arbitrary reference values $x_0$ and $\lambda_0$, and define:
\ba
\chi(E_Y; x, \lambda; x_0, \lambda_0) \equiv 
 S^0_Y(E_Y; x,\lambda)  - S^0_Y(E_Y; x_0,\lambda_0) ,
 \nonumber\\
 \label{chi-def}
\ea
 Now the crucial observation is that whilst $S^0_Y(E_Y; x,\lambda)$ is extensive in Y, Eq.~(\ref{chi-def}), the change $S^0_Y$ due to variation of $x$ and $\lambda$ is, subextensive in Y.  This is because $\lambda$ and $x$, according to our assumption, does not affect the bulk property of Y.  Hence the physical change induced by changes of $\lambda$ and $x$ must be localized around the system.   If the reader is confused, he/she should ask how the entropy of a solvent changes as a colloid moves inside it, or as we tune the interfial properties of the colloid.  

Let us now define a new Hamiltonian for Y:
\ba
H_Y(y; x,\lambda,E_Y) = H^0_Y(y; x,\lambda) 
+ T\, \chi(E_Y, x; \lambda; x_0, \lambda_0), 
 \nonumber\\
\label{H_Y-AIH-def}
\ea
where $T$ is the temperature of $Y$:
\ba
\frac{1}{T} = \beta \equiv \frac{\partial S^0_Y(E_Y; x,\lambda) }{\partial E_Y}.
\label{T-def-1}
\ea
Note that $H_Y(y; x,\lambda,E_Y)$ differs from $H^0_Y(y; x,\lambda)$ by a $y$-independent and sub-extensive constant.  
Now for exactly the same reason that Eq.~(\ref{chi-def}) is subextensive in Y, the temperature defined by Eq.~(\ref{T-def-1}) is independent of $x$ and $\lambda$---the temperature $T$ is a bulk property of Y and is insensitive to the variations of system state $x$ and of control parameter $\lambda$. 

The Boltzmann entropy of $Y$ with respect to the new bath Hamiltonian (\ref{H_Y-AIH-def}) is given by 
\ba
S_Y(E_Y; x,\lambda) &=& \log \Omega_Y(E_Y; x,\lambda)
\label{Omega_Y-new-1-1}
\\
&=& \log \int dy\, \delta (E_Y - H_Y(y; x,\lambda))
\nonumber\\
&=& \log \int dy\, \delta (E_Y -\chi  - H^0_Y(y; x,\lambda)) 
\nonumber\\
&=& \log \Omega^0_Y(E_Y -\chi )
= S^0_Y(E_Y -\chi ). \nonumber
\ea
But $\chi$ is subextensive in $Y$ and hence much smaller than $E_Y$.  We can expand $S^0_Y(E_Y -\chi )$ in terms of $\chi$ to the first order.  Further using Eqs.~(\ref{T-def-1}) and (\ref {chi-def}), we find
\ba
S_Y(E_Y; x,\lambda) &=&
 S^0_Y(E_Y; x,\lambda) - \frac{\partial 
 S^0_Y(E_Y; x,\lambda)}{\partial E_Y} \chi
\nonumber\\
&=&  S^0_Y(E_Y; x,\lambda) - \beta \chi 
 \nonumber\\
&=&  S^0_Y(E_Y; x_0,\lambda_0), 
\label{S-S_0}
\ea
from which Eqs.~(\ref{dE_Y-eq-0}) immediately follow.  Hence Eq.~(\ref{H_Y-AIH-def}) is indeed an adiabatically invariant Hamiltonian for the bath. 

The function  $\chi$, defined by Eq.~(\ref{chi-def}), characterized the local change of physics as $x$ and $\lambda$ are tuned.  It therefore can depend on $E_Y$ only through the bath temperature.  Consequently, $H_Y$, defined by Eq.~(\ref{H_Y-AIH-def}) depends on $E_Y$ only through the temperature.  We shall therefore write it as $H_Y(y; x,\lambda, \beta)$.   As long as the variation of temperature is negligible, we can further write $H_Y$ as $H_Y(y; x,\lambda)$.  

Therefore, $H_Y$ and $H_Y^0$, related by Eq.~(\ref{H_Y-AIH-def}), defines two distinct microcanonical ensembles, with distinct Boltzmann entropy, which are Eqs.~(\ref{S_Y}) and (\ref{S^0_Y}) respectively.   Careful readers may ask the following question: Which microcanonical ensemble is physical?   The answer is simple and interesting: both microcanoncal ensembles are equivalent as far as the bulk properties of the bath are concerned.  As we have already seen from Eq.~(\ref{S-S_0}), two Boltzmann entropies differ from each other only by a subextensive term.  Just like tuning of the colloidal property does not influence the bulk proerties of the ambient fluid, tuning of the function $\chi$ in Eq.~(\ref{H_Y-AIH-def}) does not change the bulk properties of the microcanonical ensemble of the bath.   As far as the bulk properties of the fluid are concerned, Eq.~(\ref{H_Y-AIH-def}) behaves much like a gauge transform.

\subsection{Hamiltonian of Mean Force}

Let us now decompose the total Hamiltonian (\ref{H_tot-decomp-0}) in the following form:
\ba
H_{\text{tot}}(x, y; \lambda) = H_X(x; \lambda) + H_Y(y; x,\lambda),
\label{H-decomp-1}
\ea
where $H_Y(y; x,\lambda)$ is given by Eq.~(\ref{H_Y-AIH-def}).  Because the total Hamilonian is independent of $\beta$, the $\beta$ dependence of $H_Y$ also demands $\beta$ dependence of $H_X$. 

We now assume that the the joint system achieves thermal equilibrium with a super-bath with temperature $T$.  The equilibrium pdf of the joint system is then 
\ba
p^{\rm eq}(x,y) = \frac{1}{Z_{XY}(\beta, \lambda)} e^{- \beta H_X(x; \lambda) - \beta H_Y(y; x; \lambda)}. 
\ea
 whereas the marginal equilibrium pdf of $x$ is:
\ba
p^{\rm eq}(x) =  \frac{1}{Z_{XY}} e^{- \beta H_X(x; \lambda)} 
\int dy\, e^{ - \beta H_Y(y; x; \lambda)}. 
\label{p_eq-x-1}
\ea
But the integral in the r.h.s. is precisely the canonical partition function of $Y$ (conditioned on $x$ and $\lambda$) and is related to the microcanonical partition function (\ref{Omega-def}) via a  Laplace transform: 
\ba
Z_Y = \int dy\, e^{ - \beta H_Y(y; x; \lambda)}
= \int e^{-\beta E_Y} \Omega(E_Y). 
\ea
Since $\Omega(E_Y)$ is independent of $x$ and $\lambda$, so is $Z_Y(\beta)$:
\ba
\frac{\partial Z _Y}{\partial \lambda}  = \frac{\partial Z _Y}{\partial x} = 0. 
\label{dZ_Y-eq-0}
\ea  
Therefore  Eq.~(\ref{p_eq-x-1}) becomes
\ba
p^{\rm eq}(x,y) = \frac{Z_{XY}}{Z_{XY}} e^{- \beta H_X(x; \lambda)},
\ea
which has a Gibbssian form with $H_X(x; \lambda)$ playing the role of an effective Hamiltonian.  Therefore $H_X(x; \lambda)$ is \emph{ the Hamiltonian of mean force} of the system.  The decomposition of Hamiltonian (\ref{H-decomp-1}) is therefore fully consistent with our earlier theories of strong coupling stochastic thermodynamics~\cite{Xing-SC-1,Xing-SC-2}.

\section{Coarse-grained Joint Dynamics and Generalized Detailed Balance}

\subsection{Microscopic Joint Dynamics}
\label{sec:micro-evo}

We note $c^*$ to denote the time-reversal of $c$, whether $c$ is a control parameter or a state variable.  Let $a, b$ be respectively the even and odd components of $c$, we then have $c= (a,b)$ and $c^* = (a, -b)$.   We assume that the system Hamiltonian has the following time-reversal symmetry:
\ba
H_X(x; \lambda) &=& H_X(x^*; \lambda^*),
\label{symm-H_X}\\
H_Y(y; x, \lambda) &=& H_Y(y^*; x^*, \lambda^*),
\label{symm-H_Y}
\ea

To simplify notations, we shall use $z = (x, y)$ to denote the collection of system and bath variables.  The microscopic joint dynamics is deterministic and governed by the Hamiltonian $H(z; \lambda) = H_{\rm tot}(z; \lambda)$ [c.f. Eq.~(\ref{H_tot-decomp})] with a time-dependent control parameter, as well as a non-conservative force.   The concrete form of Hamiltonian equations are discussed in Sec.~\ref{sec:W-Q}.   We sahall not need these details in this section.  In the forward process, the control parameter changes with the running time  $s$ as $\lambda_s$, with initial value $\lambda_0$ and final value $\lambda_t$.  In the backward process, the control parameter changes as $\tilde \lambda_s = \lambda^*_{t - s}$, with initial value $\lambda_t^*$ and final value $\lambda_0^*$.  Note that in both the forward process and the backward process, the time runs from $0$ to $t$.   We may assume that the conservative force also depends on $\lambda$, so that its functional forms are $f(z, \lambda_s)$ and $f(z, \tilde \lambda_s)$  in the forward and backward processes respectively.  These defines the dynamic protocols of the forward and backward processes.  The initial states of these processes will be discussed in Sec.~\ref{sec:CG-DB-J}.


 Let $z_0$ be an arbitrary initial state, which is carried to $z_t$ by the forward dynamics.  The relation between $z_t$ and $z_0$ defines the forward evolution operator $U(t,0)$ (the Liouville operator): 
 \ba
 z_t = U(t,0) z_0, \label{z-U-z}
 \ea 
 which is generally a nonlinear invertible operator on the state space.  The corresponding evolution operator of the backward process is denoted as $\tilde U(t,0)$.  \emph{The time-reversal symmetry} of the microscopic dynamics then tells us 
 \ba
 z_0^* = \tilde U(t,0) z_t^*.  \label{z-tilde-U-z}
 \ea
 In other words, if the forward dynamics transforms $z_0$ to $z_t$, then the backward dynamics transforms $z_t^*$ to $z_0^*$.  In the below, we shall use simplified notations $U = U(t,0)$ and $\tilde U = \tilde U(t,0)$, as there is no danger of confusion. Combining the preceding two equations, we find
\ba
 \tilde U (U z_0)^* = z_0^*.  \label{UUzz}
\ea
   
  Now let $A$ be a set, e.g. a Boltzmann cell, in the state space of $z$.  We define $UA$ as the image of $A$ created by $U$:
 \ba
 U A \equiv \{Uz \mid z \in A\}. 
 \ea
Basic set theory then tells us $U (A\cap B) =  U A \cap UB$ and $U(A \cup B) = U  A \cup U B$.   We use $A^*$ to denote the time-reversal of $A$: $A^* \equiv \{x^* \mid x \in A \}$.  From Eq.~(\ref{UUzz}) we have
  \ba
  \tilde U (U A)^* = A^*.  \label{UUAA-1}
 \ea 
 Since $A$ is arbitrary, we can also write Eq.~(\ref{UUAA-1}) as
 \ba
   \tilde U A^*  = (U^{-1} A)^*.  \label{UUAA-2}
 \ea
 We further define $|A|$ as the volume of set $A$.  Then the time-reversal symmetry of the state space tells us volume is invariant under time-reversal:
 \ba
 |A^*| = |A|. \label{A^*A}
 \ea
 whereas  the Liouville theorem tells us that volume is conserved by the dynamics:
 \ba
|U A| = |\tilde U A| = |A|.     \label{Liouville}
 \ea
 
 \subsection{Coarse-Graining and Generalized Detailed Balance}
  \label{sec:CG-DB-J}
 
We partition the  \( X  Y \)  phase space into tiny Boltzmann cells of equal volume, say, $d^n z$.   We use $A(z)$ to denote the cell centered at $z$.  From now on therefore $z$ is discretized, and each $z$ labels a Boltzmann cell.  We choose the partitioning of phase space to preserve the time-reversal symmetry, which means 
\ba
A(z^*)  = A^*(z) , \quad
 |A(z^*)|  = |A^*(z)| = |A(z)|.  
 \label{A*A*}
\ea

Let us now define the initial state of the forward process.  We assume that the system starts from a probability distribution function that is constant inside a particular Boltzmann cell $A(z)$ and vanishes everyelse.  Such a pdf is invariant under coarse-graining, i.e. local averaging inside every Boltzmann cell.  This is a special case of Boltzmann macro-state widely studied in statistical mechanics, where all micro-states inside a cell are considered indistinguishable and thereare assigned equal probability density.   More general initial pdf  can be constructed as convex superposition of these Boltzmann states.   We then let the system evolve according to the microscopic evolution operator discussed in Sec.~\ref{sec:micro-evo}.   To achieve an intuitive understanding, it is better to discuss the evolution of the Boltzmann cell, rather that that of the pdf.  At the final time $t$, the cell $A(z)$ is evolved into $U A(z)$, which, due to the ergodicity and mixing properties of the dynamics, is very complex and has overlap with a very large number of cells~\footnote{In the special case of fixed control parameter and vanishing non-conservative force, the system dynamics is conserved and stationary. Then  as $t \rightarrow \infty$, $U(t, 0) A(z)$ stays in the same energy shell, and eventually spread uniformly, in the coarse-grained sense, to the entire energy shell.  The coarse-grained pdf then approaches the equal probability distribution in the energy shell, as postulated by Boltzmann.}.     

In the coarse-grained joint dynamics, we only care about the integrated probability inside each Boltzmann cell, and do not care about the varations of the pdf inside each cell.  In other words, we focus on the coarse-graining of the pdf.  \emph{The transition probability} $P_F(z' |z)$ is defined as the probability that the system is inside the Boltzmann cell $A(z')$ at time $t$, given that it starts from a uniform pdf inside $A(z)$ at time $0$.   According to our discussion above, this is the volume fraction of the subset of $A(z)$ that is evolved into $A(z')$ by $U(t,0)$.  But the subset of $A(z)$ that is evolved into $A(z')$ is $U ^{-1}A(z') \cap A(z)$. 
Hence the forward transition probability $P_F(z' |z)$ is given by 
\ba
P_F(z' |z) = \frac{|U ^{-1}A(z') \cap A(z)|}{|A(z)|}.
\label{P_F-1-1}
\ea
Similarly, the backward transition probability is
\ba
P_B(z^* |z'^*) 
=  \frac{|A(z^*) \cap \tilde U A(z'^*)|}{|A(z'^*)|}.
\ea   

Now using Eqs.~(\ref{A*A*}), (\ref{UUAA-2}), and (\ref{A^*A}) in order, we obtain
\ba
P_B(z^* |z'^*) 
&=&  \frac{|A(z)^* \cap \tilde U A(z')^*|}{|A(z')^*|} 
\nonumber\\
&= & \frac{|A(z)^* \cap  (U^{-1} A(z'))^*|}{|A(z')^*|} 
\nonumber\\
&= & \frac{|A(z) \cap  (U^{-1} A(z'))|}{|A(z')|}. 
\ea
Finally using Eq.~(\ref{P_F-1-1}) we find
\ba
P_F(z' |z) = P_B(z^* |z'^*),
\ea
which can be expressed in terms of slow and fast variables $x, y$ as 
\ba
P_F(x', y' | x, y) = P_B(x^*, y^* | x'^*, y'^*). 
\label{P_F-sym}
\ea
Equation~\eqref{P_F-sym} is not a conventional form of DB, since it is derived for non-stationary driven systems.  Neither is it a form of LDB, since it does not involve thermodynamic quantities such as heat or entropy change.  It expresses a time reversal symmetry of the coarse-grained dynamics of a dynamically driven and thermally isolated system.  Most importantly, this dynamic symmetry remains valid even if the system is very far away from equilibrium.  It constitutes a generalization of detailed balance to far-from-equilibrium settings.  We will therefore refer to this relation as \emph{the generalized detailed balance (GDB)}~\footnote{Note that our definition of GDB is different from that in Sec. 4.3 of Ref.~\cite{Hanggi-Stochastic-review-1982}.}.  we shall use it as the starting point for deriving local detailed balance for the reduced system dynamics.  In Sec.~\ref{sec:LDB}, we will show that, when we integrate out the fast variables $y$ from Eq.~(\ref{P_F-sym}), we obtain the condtion of LDB (\ref{P_F-sym}), which supplies the essential link betwee non-equilibrium kinetics and non-equilibrium thermodynamics.

\section{Markovian Property of the Reduced Dynamics}
\label{sec:Markov}

In general, the coarse-grained joint dynamics as we constructed in Sec.~\ref{sec:CG-DB-J} is not Markovian.  Suppose the system starts from a uniform pdf in the cell $A(z_0)$, and we measure the system at time $t$ and find in the cell $z$.  The pdf of the system variable at time $t$, conditioned on the measurement result, is localized in the cell $A(z)$, but is not uniform in $A(z)$.   The non-uniform details of the pdf inside the cell $A(z)$ in general has impact on the future coarse-grained evolution of the system.   We {\em assume} that for times longer than the bath mixing time $\tau_{\rm MC}$, the coarse-grained joint dynamics effectively loses memory of subcell structure, thus becoming {\em Markovian}. Mathematically, Markovianity of the coarse-grained joint dynamics means that if for an intermediate time $s$ such that $s \gg \tau_{\rm MC}$ and $t - s \gg \tau_{\rm MC}$, we have  the following Chapman-Kolmogorov equation:
\ba
&& P_F(x_t, y_t | x_0, y_0)
 \nonumber\\
 &=& \int d x_s \int d y_s \, 
 P_F(x_t, y_t | x_s, y_s)  P_F(x_s, y_s | x_0, y_0).  \quad\quad\quad
 \label{Chapman-Kolmogorov}
\ea
Intuitively, Markovianity means that it does not matter whether how the joint system evolves before time $t$,  as long as we find that it is in cell $A(z)$ at $t$, we can use the uniform pdf is uniform inside $A(z)$ as a new initial state and study the future evolution.  This guarantees that, as long as we discretize  the time-axis with a step-time longer than the mesoscopic time $\tau_{\rm MC}$, given the state $z$ at time $t$, the state $z'$ at a later time $t'> t$ and the state $z''$ at an earlier time $t''< t$ are independent of each other, which is a defining property of Markov process. This Markovian postulate is equivalent to assuming the certain kind of mixing properties of the joint \( X Y \) dynamics.  The exact mathematical conditions that guarantee the Markovian nature of the coarse-grained dynamics is currently not understood.  Nonetheless, both experimental evidences and physical intuition indicate that Markovian dynamics can be a good approximation to most common Hamiltonian systems, as long as the coarse-graining is carried properly.  We further assume that the scheme of coarse-graining preserves both energy conservation and time-reversal symmetry.  

Suppose the coarse-grained joint system start from $(x_0, y_0)$ at $t = 0$ and evolves to $(x_t, y_t)$ at time $t$.  (Recall that both $(x_0, y_0)$ and $(x_t, y_t)$  denote Boltzmann cells.)  According to the assumption of TSS, $y_t$ is in equilibrium conditioned on $x_t$, which means that $P_F(x_t, y_t | x_0, y_0)$ may depend on $y_t$ only via the final bath energy $H_Y(y_t;x_t, \lambda_t)$.  For any initial state $y_0$, at time scale $ \tau_y \ll \tau_x$, the bath \( Y \) equilibrates microcanonically, whereas the system variables and the control parameter barely changed.  This implies that $P_F(x_t, y_t | x_0, y_0)$ may depend on $y_0$ only via the initial bath energy $H_Y(y_0;x_0, \lambda_0)$.  Therefore we can write $P_F(x_t, y_t | x_0, y_0)$, the transition probability of the forward process, in the following form:
\ba
&& P_F(x_t, y_t | x_0, y_0)
 \nonumber\\
&=  &
\left\langle x_t, H_Y(y_t;x_t, \lambda_t) |  x_0, H_Y(y_0;x_0, \lambda_0) \right\rangle_F,
\label{P_F-XY-interm}
\ea
where $\left\langle x', E_Y' |  x, E_Y\right\rangle_F$ can be understood as the probability density that the joint system starts from state $(x, y)$, with $E_Y = H_Y(y;x, \lambda)$, and transits to state $(x', y')$, with $E_Y' = H_Y(y';x', \lambda')$. Similarly for the backward process we similarly have:
\ba
&& P_B(x_0^*, y_0^* | x_t^*, y_t^*)
\nonumber\\
 &=& 
\left\langle x_0^*, H_Y(y_0^*;x_0^*, \lambda_0^*) |
  x_t^*, H_Y(y_t^*;x_t^*, \lambda_t^*)  \right\rangle_B 
  \nonumber\\
 &=& 
\left\langle x_0^*, H_Y(y_0;x_0, \lambda_0) |
  x_t^*, H_Y(y_t;x_t, \lambda_t)  \right\rangle_B,
\ea
where used was the symmetry (\ref{symm-H_Y}).  The GDB condition (\ref{P_F-sym}) can then be rewritten as
\ba
\left\langle x', E_Y' |  x, E_Y \right\rangle_F
= \left\langle x^*, E_Y |
  x'^*, E_Y' \right\rangle_B. 
\label{P_F-sym-1}
\ea


For convenience, we intrdouce the following shorthands for the initial and final energies for $X$ and $Y$:
\ba
H_{X,0} \equiv H_X(x_0; \lambda_0), &\,\,&
H_{Y,0} \equiv H_Y(y_0;x_0, \lambda_0), \quad \\
H_{X,t} \equiv H_X(x_t; \lambda_t), &\,\,&
H_{Y,t} \equiv H_Y(y_t;x_t, \lambda_t). 
\ea
We integrate Eq.~(\ref{P_F-XY-interm}) over the final  bath state $y_t$, and obtain the marginal pdf for $x_t$:
\ba
P_F(x_t | x_0, y_0) 
&\equiv& 
\int dy_t \, P_F(x_t, y_t | x_0, y_0)
\nonumber\\
&=& \int dy_t \,  \left\langle x_t, H_{Y,t} |  
x_0, H_{Y,0} \right\rangle_F. \quad \label{P_F-X-int-1}
\ea
Inserting a Dirac delta function, we find
\ba
&&P_F(x_t | x_0, y_0) 
\nonumber\\
&=&  \int dy_t \int dE_Y \delta(H_Y - E_Y)\,
 \left\langle x_t, E_Y |  
x_0, H_{Y,0} \right\rangle_F
\nonumber\\
&=&  \int dE_Y \Omega_Y(E_Y)  \left\langle x_t, E_Y |  
x_0, H_{Y,0} \right\rangle_F,
\nonumber\\
&=&  \int dE_Y e^{S_Y(E_Y)}  \left\langle x_t, E_Y |  
x_0, H_{Y,0} \right\rangle_F,
\label{P_F-X-int-2}
\ea
where $ \Omega_Y(E_Y) , S_Y(E_Y)$ are independent of $x; \lambda$, as ensured by the adiabatic invariance of $H_Y$. 

Equation (\ref{P_F-X-int-2}) is the probability density that the system ends at $x_t$, given that the system starts from $x_0$ and the bath starts with energy $H_{Y,0} $.   In other words,  it is \emph{ the transition probability of the reduced $X$ dynamics}, conditioned on the initial energy of the bath.  Now a crucial observation is that the initial bath energy can influence the reduced $X$ dynamics only through the temperature.  This is enforced by the locality of interaction and the extensive nature of the bath.  Suppose we add to the bath an energy $\delta E$ which is much smaller than the bath energy $H_{Y,0} $.  The resulting change of the bath temperature is negligible, and hence does not affect the reduced $X$ dynamics. (Think about adding one Joule of heat to the pacific ocean and ask how it influences to the swimming of a fish inside.)   We can therefore rewrite Eq.~(\ref{P_F-X-int-2}) as
\ba
P_F(x_t | x_0; \beta ) 
\equiv 
\int dE_Y e^{S_Y(E_Y)}   \left\langle x_t, E_{Y,t} |  
x_0, H_{Y,0} ) \right\rangle_F, 
\nonumber\\
\label{P_F-X-int-3}
\ea
and interpret it as the transition probability of the reduced $X$ dynamics at a given bath temperature.  \emph{This is the physical essence of time-scale separation:  in the limit of large bath size, only the intensive parameters of the bath affect the system dynamics. }   The integrant $e^{S_Y(E_{Y,t})}  \left\langle x_t, E_{Y,t} |  x_0, E_{Y,0} \right\rangle_F$ can then be understood as the joint probabilty density that, at time $t$, the system is in state $x_t$ and the bath has energy $E_{Y,t}$, given that, at time $0$, the system is in state $x_0$ and the initial bath energy is  $E_{Y,0}$.

Recall that the joint $XY$ dynamics is Markovian. 
Integrating both sides of Eq.~(\ref{Chapman-Kolmogorov}) over $y_t$, and using Eq.~(\ref{P_F-X-int-1})-(\ref{P_F-X-int-3}), we find
\ba
&& P_F(x_t | x_0; \beta )
 \nonumber\\
 &=&
 \int d x_s \int  d y_s  P_F(x_t | x_s; \beta ) 
 P_F(x_s, y_s | x_0, y_0).  \quad\quad
 \label{Chapman-Kolmogorov-2}
\ea
Since $P_F(x_t | x_s; \beta ) $ is independent of $y_s$, we can pass the $y_s$ integral through and rewrite Eq.~(\ref{Chapman-Kolmogorov-2}) as
\ba
P_F(x_t | x_0; \beta ) \!\! &=& \!\!
 \int d x_s P_F(x_t | x_s; \beta ) 
\int  d y_s    P_F(x_s, y_s | x_0, y_0).  
\nonumber\\
& =& \int d x_s P_F(x_t | x_s; \beta )
 P_F(x_s | x_0; \beta ),
 \label{Chapman-Kolmogorov-3}
 \ea
which is precisely the Chapman-Kolmogorov equation of the reduced $X$ dynamics.  It guarantees that the reduced $X$ dynamics is also Markovian. 

Conventional projection operator techniques, such as the Zwanzig--Mori formalism, were used to establish Markovian property by assuming weak interactions and near equilibrium. Our previously developed multi-scale projection operator method~\cite{Xing-CG}, while free from the weak coupling and near equilibrium constraints, still relies on a Langevin description of the system dynamics.  By contrast, the present work establishes the Markovian nature of the reduced dynamics directly from TSS, without appealing to any specific dynamical details. This makes our proof of Markovianity both more general and more fundamental.

\section{Work and Heat}
\label{sec:W-Q}

Let $\tau_x$ and $\tau_y$ be respectively the time scales of the slow and fast variables. According to  \emph{ TSS},  we can find an intermediate time scale $t$ with $ \tau_y \ll t \ll \tau_x$, such that \( Y \) equilibrates microcanonically conditioned on \( X \), while \( X \) remains non-equilibrium. The external parameter \( \lambda_t \) evolves from \( \lambda_0 \) at \( t=0 \) to \( \lambda_t \) at time \( t \), assumed continuously and slowly varying relative to \( \tau_y \).  

The joint system $XY$ is isolated and evolves according to the Hamiltonian dynamics. Let the system starts at \( (x_0, y_0) \) with total energy:
\ba
E_0^{\text{tot}} = H_X(x_0, \lambda_0) + H_Y(y_0; x_0, \lambda_0),
\ea
and ends at \( (x_t, y_t) \) with:
\ba
E_t^{\text{tot}} = H_X(x_t, \lambda_t) + H_Y(y_t;x_t, \lambda_t).
\ea

The work done by the external agent controlling \( \lambda_t \) is the change in total energy:
\ba
W &=& E_t^{\text{tot}} - E_0^{\text{tot}} 
 \nonumber\\
&=& H_{\text{tot}}(x_t, y_t, \lambda_t) - H_{\text{tot}}(x_0, y_0, \lambda_0)
\label{W-E_tot-diff}\\ 
&=& [H_X(x_t, \lambda_t) + H_Y(y_t;x_t, \lambda_t)]
 \nonumber\\
 &-& [H_X(x_0, \lambda_0) + H_Y(y_0;x_0, \lambda_0)].
\nonumber
\ea
This form seems to suggest that the work depends both on $x_t$ and $y_t$, as well as the protocol $\lambda_t$.  This is in fact not true.  To see it, we note that the work can be rewritten as a time integral:
\ba
 W = H_{\text{tot}}(x_t, y_t, \lambda_t) - H_{\text{tot}}(x_0, y_0, \lambda_0)
 =  \int_0^t  \frac{d H_{\text{tot}}}{ds} ds,
 \label{W-int-1}
\ea
where $s$ is the running time.  Let us compute the total derivative:
\ba
\frac{d H_{\text{tot}}}{ds} = \frac{\partial H_{\text{tot}}}{\partial x} \frac{dx}{ds} 
+ \frac{\partial H_{\text{tot}}}{\partial y} \frac{dy}{ds} 
+ \frac{\partial H_{\text{tot}}}{\partial \lambda} \dot{\lambda}(s).
\label{dH-tot-dt}
\ea

We consider a Hamiltonian system that is driven both by a time-dependent control parameter and by a generally nonconservative external force.  (The processes we study are therefore slightly more general that those conventionally studied in strong coupling stochastic thermodynamics.)  Let $x = (q_x, p_x)$, $y= (q_y, p_y)$ by the coordinate and momentum of the system and the bath. Their evolutions obey the Hamiltonian equations:
\begin{subequations}
\ba
\frac{d q_x}{ds} = \frac{\partial H_{\rm tot}}{\partial p_x}, &\quad&
\frac{d p_x}{ds} = -  \frac{\partial H_{\rm tot}}{\partial q_x}
+ f(q_x), \\
\frac{d q_y}{ds} = \frac{\partial H_{\rm tot}}{\partial p_y}, &\quad&
\frac{d p_y}{ds} = -  \frac{\partial H_{\rm tot}}{\partial q_y},
\ea
\label{Hamiltonian-eqns}
\end{subequations}
where $f(q_x)$ is the (generally non-conservative) driving force acting on the system coordinate.  We assume that $f$ does not depend on system momentum, so that the Liouville theorem remains hold. 

Using Eqs.~(\ref{Hamiltonian-eqns}), we easily compute:
\ba
\frac{\partial H_{\text{tot}}}{\partial y} \frac{dy}{ds}  
&=& \frac{\partial H_{\text{tot}}}{\partial q_y} \frac{dq_y}{ds}  
+ \frac{\partial H_{\text{tot}}}{\partial p_y} \frac{dp_y}{ds}  
= 0\\
\frac{\partial H_{\text{tot}}}{\partial x} \frac{dx}{ds}  
&=& \frac{\partial H_{\text{tot}}}{\partial q_x} \frac{dq_x}{ds}  
+ \frac{\partial H_{\text{tot}}}{\partial p_x} \frac{dp_x}{ds}  
= f \frac{d q_x}{ds} .  \quad
\ea
Consequently we can rewrite Eq.~(\ref{dH-tot-dt}) as
\ba
{d H_{\text{tot}}} =  \frac{\partial H_{\rm tot}}{\partial \lambda} d\lambda
+ f \, d q_x.
\ea 

Equation (\ref{W-int-1}) then reduces to 
\ba
 W = E_t^{\text{tot}} - E_0^{\text{tot}}
=  \int_0^t   f \, d q_x(s) 
+ \int_0^t  \frac{\partial H_{\rm tot}(x, y; \lambda)}{\partial \lambda}
 \, d\lambda(s).
 \nonumber\\
\label{W-int-x-1} 
\ea
The integral due to variation of $\lambda$ in the r.h.s. can be decomposed into:
\ba
\int_0^t  \frac{\partial H_{\rm tot}(x,y; \lambda)}{\partial \lambda} \, d\lambda
&=& \int_0^t  \frac{\partial H_{X}(x; \lambda)}{\partial \lambda} \, d\lambda
 \nonumber\\
&+& \int_0^t  \frac{\partial H_{Y}(y; x; \lambda)}{\partial \lambda} \, d\lambda.
\quad
\ea
The integral involving $H_y$ is typically oscillating rapidly with time-scale $\tau_y$, and must be interpreted as the work done to the bath.  In the time-scale $\tau_x$, Y equilibrates, hence the time-average is identical to the microcanonical ensemble average [c.f. Eq.~(\ref{p_Y-cond-EQ})]:
\ba
 \int_0^t  \frac{\partial H_{Y}(y; x; \lambda)}{\partial \lambda} \, d\lambda
 &\approx& \int dy\, \frac{\partial H_{Y}(y; x; \lambda)}{\partial \lambda} 
 \frac{ \delta (H_Y - E_Y) }{\Omega_Y}
\nonumber\\
&=& \int_0^t  \frac{\partial \log \Omega_Y}{\partial \lambda} d \lambda
= 0,
\ea
where in the last equality we have used Eq.~(\ref{Omega_Y-eq-0-2}).  Therefore the work done by $\lambda$ to the bath averages out in the slow time-scales.  This is of course consistent with our discussion in Sec.~\ref{sec:decomp-H}. 


Since we are only interested in the dynamics of the system, we shall ignore these rapid fluctuations of work in Eq.~(\ref{W-int-x-1}), and rewrite the work as
\ba
 W =  
\int_0^t  \frac{\partial H_X(x; \lambda)}{\partial \lambda} \, d\lambda(s)
+ \int_0^t   f \, d q_x(s) ,
\label{W-int-x-2} 
\ea
which agrees with the result derived in Ref.~\cite{Xing-NC-ST}.  As one can see, there is no need to know the detailed trajectory of the bath variables in order to compute the work. 



Recall that the heat \( Q \) is negative the change of the bath energy [c.f. Eq.~(\ref{dE-S-Q-Y})]:
\ba
Q = - \left( H_Y(y_t;x_t, \lambda_t) - H_Y(y_0;x_0,\lambda_0) \right).
\label{Q-def-H}
\ea
Combining this with Eqs.~(\ref{W-E_tot-diff}) and (\ref{H-decomp-1}), we see 
\ba
W = - Q + H_X(x_t; \lambda_t) - H_X(x_0; \lambda_0),
 \label{first-law-0}
 \ea
which can be further rewritten as
\ba
\Delta E_X =  H_X(x_t; \lambda_t) - H_X(x_0; \lambda_0)
= W + Q, \label{first-law-1}
\ea
which is interpreted as the first law of thermodynamics at the trajectory level.  From this we further obtain the following expression for heat:
\ba
Q =  \int_0^t   \frac{\partial H_X}{\partial x} dx  
-  \int_0^t   f \, d q_x,
\label{Q-int-1}
\ea
which can also be computed using the trajectroy of the system variable alone. 

For an equilibrium system, there is no driving force, $f = 0$, and the control parameter remains fixed. Consequently, the work (\ref{W-int-x-2}) vanishes identically, and the heat (\ref{Q-int-1}) becomes the integral of the full differential of system Hamiltonian:
\ba
Q = \int_0^t dH_X = H_X(x_t) - H_X(x_0). 
\label{Q-int-eq}
\ea

If we use Langevin dynamics with white noises to model the system dynamics,  the expressions (\ref{W-int-x-2}) and (\ref{Q-int-1}) remain valid in the Markov dynamics, as long as we interpret various products in Stratonovich's sense. 
In the absence of non-conservative forces, these definitions of work and heat are fully consistent with our earlier theories in Refs.~\cite{Xing-SC-1} and \cite{Xing-SC-2}.

\section{Local Detailed Balance}
\label{sec:LDB}

According to the discussion after Eq.~(\ref{P_F-X-int-3}), we may understand $e^{S_Y(E_{Y,t})}  \left\langle x_t, E_{Y,t} |  x_0, E_{Y,0} \right\rangle_F$ as the joint probabilty density that, at time $t$, the system is in state $x_t$ and the bath has energy $E_{Y,t}$, given that, at time $0$, the system is in state $x_0$ and the initial bath energy is  $E_{Y,0}$.   In such a process, the heat released by the bath is 
\ba
Q = E_{Y,0}  - E_{Y,t}. \label{Q-t-0-1}
\ea
We can then rewrite $e^{S_Y(E_{Y,t})}  \left\langle x_t, E_{Y,t} |  x_0, E_{Y,0} \right\rangle_F$ as 
\ba
P_F(Q, x_t | x_0;\beta )  =  e^{S_Y(E_{Y,0} - Q) } 
 \left\langle x_t, E_{Y,0} - Q |  
x_0,   E_{Y,0}\right\rangle_F,
\nonumber\\
\label{P_F-Q-x}
\ea
and interpret it as the joint pdf of the heat $Q$ and the final state $x_t$.   The notation in the l.h.s. of Eq.~(\ref{P_F-Q-x}) indicates that  this joint pdf depends on $E_{Y,0}$ only via $\beta$.   We can easily verify that integration of Eq.~(\ref{P_F-Q-x}) over $Q$ yields  the marginal pdf for $x_t$:  
\ba
P_F(x_t | x_0; \beta )  = \int dQ\, P_F(Q,x_t | x_0; \beta ).
\ea

We can now construct the analogue of Eq.~(\ref{P_F-Q-x}) for the backward process, where the control parameter evolves as $\lambda^*_{t - s}$, and the system evolves from $x_t^*$ to $x_0^*$, the bath energy starts with $E_{Y,t} = E_{Y,0} - Q$ and ends with $E_{Y,0}$.  The heat transfer of this backward process is clearly $-Q = E_{Y,t} - E_{Y,0}$. 
 The joint probability density of the final state and the heat transfer is then
\ba
P_B(- Q, x_0^* | x_t^*;\beta ) = e^{S_Y(E_{Y,0}) } 
 \left\langle  x_0^*,   E_{Y,0} |x_t^*, E_{Y,0} - Q \right\rangle_B
\nonumber\\
\label{P_B-X-int-Q}
\ea
Now taking the ratio between Eqs.~(\ref{P_F-Q-x}) and (\ref{P_B-X-int-Q}), and using the GDB (\ref{P_F-sym-1}), we arrive at
\ba
\frac{P_F(Q, x_t | x_0; \beta ) }{P_B( - Q, x_0^* | x_t^*;\beta ) } 
= e^{\Delta S_Y } ,
\label{LDB-1}
\ea 
where the exponent in the r.h.s. is \emph{ the entropy change of the bath during the forward process}:
\ba
\Delta S_Y &=&S_Y(E_{Y,t}) - S_Y(E_{Y,0})
 \nonumber\\
 &=& S_Y(E_{Y,0} - Q) - S_Y(E_{Y,0}).  
 \label{Delta-S_Y-1}
\ea
 For a typical process, the heat $Q$ is much smaller than the bath energy $E_{Y,0}$.  Hence we may expand $S_Y(E_{Y,0} - Q) $ in terms of $Q$, and obtain
 \ba
 \Delta S_Y = - \beta Q, \quad \beta = \frac{\partial S_Y}{\partial E_Y}. 
 \label{DeltaS_Y-Q}
 \ea
 Here again the adiabatic invariance of $H_Y$ plays an essential role, for otherwise, Eq.~(\ref{Delta-S_Y-1}) would depend on the variations of $x$ and $\lambda$, and Eq.~(\ref{DeltaS_Y-Q}) would no longer hold. We can therefore rewrite Eq.~(\ref{LDB-1}) into the following LDB condition:
\ba
\frac{P_F(Q, x_t | x_0; \beta ) }{P_B( - Q, x_0^* | x_t^*;\beta ) }
= e^{- \beta \, Q } = e^{ \Delta S_Y},
\label{LDB-0}
\ea 
which is precisely Eq.~(\ref{LDB-1-0}) that we advertized in Sec.~\ref{sec:intro}.

Generalization of Eq.~(\ref{LDB-0}) to the case of multiple heat baths is immediate.  Each heat bath is in microcanonical equilibrium conditioned on $x, \lambda$, with Eqs.~(\ref{Delta-S_Y-1}) and (\ref{DeltaS_Y-Q}) hold.  Without bothering writing out more details, we present the LDB for multiple baths directly:
\ba
\frac{P_F(\{Q \}, x_t | x_0; \{\beta\} ) }{P_B( - \{Q\}, x_0^* | x_t^*; \{\beta\} ) }
= e^{- \sum_k \beta_k \, Q_k } = e^{ \Delta \sum_kS_{Y_k}}, \quad
\label{LDB-0-k}
\ea 
where $\beta_k$ and $Q_k$ are respectively the temperature of $k$-th bath and the heat it releases. 

For an equilibrium system, the heat reduces to Eq.~(\ref{Q-int-eq}), which is fully determined by $x_t$ and $x_0$.  The joint pdf $P_F(Q, x_t | x_0; \beta )$ then reduces to $P_F(x_t | x_0; \beta )$, and the LDB (\ref{LDB-0}) becomes 
\ba
\frac{P_F(x_t | x_0; \beta ) }{P_B(x_0^* | x_t^*;\beta ) }
= e^{ \beta \, H_X(x_t) - \beta \, H_X(x_0) }, 
\label{DB-0}
\ea
which is the well-known DB condition. 

Our LDB condition (\ref{LDB-0})  appears very different from the usual formulation (\ref{LDB-gamma}) of local detailed balance condition, which involves the probabilities dynamic trajectories (conditioned on their initial states) in the forward and backward processes.  To derive Eq.~(\ref{LDB-gamma}) from Eq.~(\ref{LDB-0}), we first assume $\tau_y \ll t = dt \ll \tau_y$, so that typically $x_{dt} - x_0 = dx$ is infinitesimal.  This corresponds to an infinitesimal step of evolution for the system variable $x$.  We can then use the infinitesimal version of Eq.~(\ref{Q-int-1}) to rewrite the heat in Eq.~(\ref{LDB-0}) as
\ba
\dbar Q = H_X(x_0 + dx; \lambda_0) - H_X (x_0, \lambda_0)
- f d q_x.  
\label{dbar-Q-1}
\ea
It does not matter whether we set $\lambda = \lambda_0$ or $\lambda = \lambda_t$---the resulting difference is of higher order.   The key point here is that the heat (\ref{dbar-Q-1}) is completely determined by $x_0$ and $x_{dt}$, with no independent stochasticity.  Consequently, the joint pdf $P_F(Q, x_{dt} | x_0; \beta )$ reduces to the pdf of $x_t$ alone, and Eq.~(\ref{LDB-0}) reduces to 
\ba
\frac{P_F(x_{dt} | x_0; \beta ) }{P_B( x_0^* | x_{dt}^*;\beta ) }
= e^{- \beta \, \dbar Q }. 
\label{LDB-infini}
\ea
This can be deemed as the infinitesimal version of Eq.~(\ref{LDB-gamma}).   Now consider a process with duration $\tau \gg \tau_x$.  We can break the process into a large number of infinitesimal steps, each with its own local detailed balance condition in the form of Eq.~(\ref{LDB-infini}).  We can then multiply all these results and obtain the local detailed balance condition for the whole process, in the form of Eq.~(\ref{LDB-gamma}).  

To derive  Eq.~(\ref{LDB-0}) from Eq.~(\ref{LDB-gamma}), we only need to rewrite Eq.~(\ref{LDB-gamma}) as
\ba
{P_F [\gamma | \gamma_0]}= e^{- \beta Q[\gamma]}
{P_B[\tilde \gamma |\tilde \gamma_0]} , 
\label{LDB-gamma-1}
\ea
and sum over all trajectories with fixed heat transfer.


Therefore, as long as the path probabilities in Eq.~(\ref{LDB-gamma-1}) are defined properly using discretization,  our version of LDB  (\ref{LDB-0}) is an equivalent to the conventional formulation of LDB (\ref{LDB-gamma-1}).   Yet Eq.~(\ref{LDB-0}) is advantageous over Eq.~(\ref{LDB-gamma}) for several important reasons.  The conventional formulation Eq.~(\ref{LDB-gamma-1}) of LDB involves probability densities of trajectories, which are difficult to construct in concrete problems.  The main reason is that, except for very simple systems, we still lack a rigorous mathematical theory for probability distributions in infinite dimensional function space.  For nonlinear Langevin systems involving multiplicative noises, for example, there have been long-lasting disputes on the proper method to compute  probability for infinitesimal paths using stochastic differential equations. For a possible resolution, see our recent work \cite{Xing-PI-2022}. Our formulation of LDB Eq.~(\ref{LDB-0}) is free of all these subtleties.  Additionally, our LDB (\ref{LDB-0}) is formulated on precisely defined concepts of heat and bath entropy. It is not only free of the ambiguity associated with the conventional form (\ref{LDB-gamma-1}), but also experimentally verifiable, since both $Q$ and $x_t, x_0$ are directly measurable for many systems.  Finally, as we shall demonstrate in the following section, our LDB (\ref{LDB-0}) supplies a much more direct route to the proof of FTs---there is no need to sum over all paths.   

The strong coupling theory of stochastic thermodynamics by Seifert et. al.~\cite{Seifert-SC}, which is rather influential, defines heat in a different way from ours.  As a consequence, the LDB (\ref{LDB-0}) does not hold for this theory.  This problem may be rescued by the different definition of system entropy, so that the unconditional version of LDB remains hold.  Nonetheless, Seifert's theory is much more complicated than the alternative, and hence much more cubersome when applied to concrete problems.

\section{Fluctuation Theorems}
\label{sec:FTs}

To demonstrate the power of our formalism, we derive a master FT involving the joint pdf of work, heat, as well as the initial and final state of the system.  From this master FT, we further derive a set of FTs,  some of which are novel.  Unlike conventional derivations, we shall not need to invoke any potential dangerous integration over trajectories. 

Let us now consider a process where the joint system starts from a microcanonical distribution with fixed total energy $E_{XY}:$
\ba
p_{XY}^0(x_0,y_0) = \frac{\delta (H_{X,0} + H_{Y,0} - E_{XY})}
{\Omega_{XY}(E_{XY}, \lambda_0)} ,
\label{p-XY^0}
\ea
where the normalization constant is 
\ba
\Omega_{XY}(E_{XY}, \lambda_0)  &=& \iint dx dy\, 
\delta (H_{\rm tot}(x_0,y_0; \lambda_0) - E_{XY}). 
\nonumber \\
&=& \iint dE_X\ dx_0 \, \delta (H_X(x_0; \lambda_0) - E_{X}) 
\nonumber\\
&\times& \int dy\, \delta (H_Y(x_0,y_0; \lambda_0) - E_{XY} + E_X)
\nonumber\\
&=& \iint dE_X\ dx_0 \, \delta (H_X(x_0; \lambda_0) - E_{X}) 
\nonumber\\
&\times&e^{S_Y( E_{XY} -  E_X)}
\nonumber\\
&=& \iint dE_X\ dx_0 \, \delta (H_X(x_0; \lambda_0) - E_{X}) 
\nonumber\\
&\times& e^{S_Y( E_{XY})} e^{- \beta E_X}
\nonumber\\
&=& \Omega_Y( E_{XY} )  \int dx_0 \, e^{- \beta H_X}
\nonumber\\
&=& e^{S_Y(E_{XY})} e^{- \beta F(\lambda_0)},
\label{Omega_XY-1}
\ea
where we have used Eqs.~(\ref{Omega-S_Y-indep}) and have expanded $S_Y( E_{XY} -  E_X)$ in terms of $E_X$ to the first order, whilst $F(\lambda_0) $ is the free energy of the system in the initial state:
\ba
F(\lambda_0) = - T \log \int dx_0 e^{- \beta H_X(x_0; \lambda_0)}. 
\ea
Hence we see that the $\lambda_0$ dependence of $\Omega_{XY}(E_{XY}, \lambda_0) $ comes from $F(\lambda_0)$.  

The reduced pdf of $x_0$ in the initial state can be easily obtained by integrating  Eq.~(\ref{p-XY^0}) over $y_0$:
\ba
p_X^0(x_0) &=& \frac{\Omega_{Y}(E_{XY} - H_{X,0})}
{\Omega_{XY}(E_{XY}, \lambda_0)}
\nonumber\\
&=& \frac{\Omega_{Y}(E_{XY} )}{\Omega_{XY}(E_{XY}, \lambda_0)} 
e^{- \beta H_X(x_0; \lambda_0)},
\ea
Further using Eq.~(\ref{Omega_XY-1}), we find  
\ba
p_X^0(x_0) = e^{\beta F(\lambda_0) - \beta H_X(x_0; \lambda_0)}, 
\label{p_X-0-canonical}
\ea
which is properly noramlized and canonical in the initial system Hamiltonian $ H_X(x_0; \lambda_0)$.  This is rather anticipated---a system that is in equilibrium with a much larger heat bath should be in a canonical equilibrium with the temperature determined by the bath. 

As in the preceding sections, we assume that the joint system start from $x_0,y_0$ with total energy $E_{XY} = H_{X,0} + H_{Y,0}$, and evovles to $x_t, y_y$.  The work is the change of the total energy:
\ba
W &=& H_{X,t} + H_{Y,t} - E_{XY} 
\nonumber\\
 &=& H_{X,t} + H_{Y,t} - H_{X,0} - H_{Y,0}. 
 \label{W-HHHH}
\ea
Further recall that the heat is [c.f. Eq.~(\ref{Q-def-H})]:
\ba
Q = H_{Y,0} - H_{Y,t}. 
\ea
We can compute joint pdf of $x_t,x_0, W, Q$:
\ba
p_F(W, Q, x_t,x_0) \!\!&=& \!\!\!\! \iint dy_t dy_0
\left\langle x_t, H_{Y,t} |  x_0, H_{Y,0}\right\rangle_F  
\nonumber\\
&\times&\frac{\delta (H_{X,0} + H_{Y,0} - E_{XY})}
{\Omega_{XY}(E_{XY}, \lambda_0)} 
\nonumber\\
&\times&  
 \delta (H_{X,t} + H_{Y,t} - E_{XY} - W) \,
 \nonumber\\
&&  \delta (H_{Y,0} - H_{Y,t} - Q).
 \ea
The integrals over $y_0,y_t$ can be carried out similar to what we have done in Eq.~(\ref{P_F-X-int-2}):
\ba
&& p_F(W, Q, x_t,x_0) 
\nonumber\\
&=& 
 \left\langle x_t, E_{XY} + W - H_{X,t} |   x_0,  E_{XY} - H_{X,0}\right\rangle_F
\nonumber\\
&\times&\Omega_Y(E_{XY} + W - H_{X,t})
 \frac{\Omega_Y(E_{XY} - H_{X,0})}
{\Omega_{XY}(E_{XY}, \lambda_0)} 
\nonumber\\
&\times& 
 \delta (H_{X,t} - H_{X,0} - W - Q),
 \label{p_F-4}
\ea
where we have also used Eq.~(\ref{W-HHHH}).  This joint pdf is singular due to the delta function which enforces the first law of thermodynamics, i.e., $ p_F(W, Q, x_t,x_0) $ vanishes identically if $W, Q, x_t,x_0$ are such that Eq.~(\ref{first-law-1}) do not hold.  
  We can integrate out any of four variables $W, Q, x_t,x_0$ to obtain a non-singular pdf. Nonetheless, we shall keep using this singular pdf to derive a master version of FT.

Let us now do the same thing to the backward process. Using the time-reversal symmetry of Hamiltonian, we find that the joint pdf evaluated at $-W,-Q, x_0^*, x_t^*$ is
\ba
&& p_B( - W, - Q,x_0^*, x_t^*) 
\nonumber\\
&=& 
 \left\langle x_0^*,  E_{XY} - H_{X,0} |  x_t^*, E_{XY} + W - H_{X,t} \right\rangle_B
\nonumber\\
&\times& \Omega_Y(E_{XY} - H_{X,0}) 
 \frac{\Omega_Y(E_{XY} + W - H_{X,t})}
{\Omega_{XY}(E_{XY} + W, \lambda_t)} 
\nonumber\\
&\times& 
 \delta (H_{X,0} - H_{X,t}  + W + Q). 
 \label{p_B-4}
\ea

Now taking the ratio between Eqs.~(\ref{p_F-4}) and (\ref{p_B-4}), and taking advantage of the detailed balance condition (\ref{P_F-sym-1}), we find
\ba
\frac{p_F(W, Q, x_t,x_0) } {p_B( - W, - Q,x_0^*, x_t^*) }
= \frac{\Omega_{XY}(E_{XY} + W, \lambda_t)}
 {\Omega_{XY}(E_{XY}, \lambda_0)}.
 \label{FT-4-1}
\ea
Recall that  ${\Omega_{XY}(E_{XY}, \lambda_0)}$ is already computed in Eq.~(\ref{Omega_XY-1}).  Hence we have 
 \ba
 && \frac{\Omega_{XY}(E_{XY} + W, \lambda_t)}
 {\Omega_{XY}(E_{XY}, \lambda_0)} 
  \nonumber\\
  &=& e^{ S_Y(E_{XY} + W) - S_Y(E_{XY})}
 e^{- \beta F(\lambda_t) + \beta F(\lambda_0)}
  \nonumber\\
 &=& e^{\beta W - \beta F(\lambda_t) + \beta F(\lambda_0)}
   \nonumber\\
 &=& e^{\beta (W - \Delta F)},
  \ea 
  where, again, played an essential role is the independence of $S_Y$ on $\lambda$.  Substituting this back into Eq.~(\ref{FT-4-1}) we finally obtain the master FT:
\ba
{p_F(W, Q, x_t,x_0) } \,  e^{ \beta ( \Delta F - W )} 
= {p_B( - W, - Q,x_0^*, x_t^*) }.  \nonumber\\
\label{FT-master}
\ea
We refer to this result as the master FT because it encapsulates the joint statistics of work, heat, and system states. All other fluctuation relations of interest can be derived by appropriate marginalization.”

We are now free to integrate any one, two, or three variables out of Eq.~(\ref{FT-master}) and obtain FT for non-singular pdfs.  For example, if we integrate out $ x_t,x_0$, we obtain the joint FT for work and heat:
\ba
p_F(W,Q) \, e^{ \beta ( \Delta F - W )}  = p_B( - W, -Q). 
\ea
If we further integrate out $Q$, we obtain the Crooks FT:
\ba
p_F(W) \, e^{ \beta ( \Delta F - W )}  = p_B( - W). 
\ea 
If we integrate out $Q$ from Eq.~(\ref{FT-master}), we obtain:
\ba
{p_F(W, x_t,x_0) } \,  e^{ \beta ( \Delta F - W )} 
= {p_B( - W,x_0^*, x_t^*) }. 
\label{FT-3}
\ea
If we further integrate out $x_t$, we obtain
\ba
{p_F(W, x_0) } \,  e^{ \beta ( \Delta F - W )} 
= {p_B( - W,x_0^*) }. 
\label{FT-3-2}
\ea
This FT must be carefully interpreted.  In the forward process, we measure the joint pdf of the initial state and the work, whereas in the backward process, we measure the joint pdf of the final state and the work. 

Finally, let us integrate the work from Eq.~(\ref{FT-master}).  Using the first law (\ref{first-law-0}), we may simply replace $W$ in the exponent of Eq.~(\ref{FT-master}) by $H_{X,t} - H_{X,0} - Q$, and obtain
\ba
{p_F(Q, x_t,x_0) } \,  e^{ \beta ( \Delta F + Q - \Delta H_X )} 
= {p_B( - Q,x_0^*, x_t^*) }, \quad
\label{FT-Q-xx}
\ea
where $\Delta H_X  = H_{X,t}  - H_{X,0} $ is the total change of the system energy along the trajectory.  Due to the energy conservation, this FT actually encodes exactly the same information as Eq.~(\ref{FT-3}).   It appears to us that FTs (\ref{FT-3}), (\ref{FT-3-2}) have not been discussed previously.   

\section{Conclusion}
\label{sec:conclusion}

Let us clarify the connections between the present work and several of our previous contributions to stochastic thermodynamics.

The decomposition of the total Hamiltonian into an adiabatically invariant bath Hamiltonian and a Hamiltonian of mean force for the system is fully consistent with our earlier works~\cite{Xing-SC-1, Xing-SC-2} on the thermodynamics of strongly coupled systems. The definitions of work and heat adopted here also agree with those proposed in our previous formulations of stochastic thermodynamics: for systems without non-conservative forces~\cite{Xing-ST-cov}, and for systems driven by non-conservative forces~\cite{Xing-NC-ST}.

The local detailed balance (LDB) condition (\ref{LDB-gamma}) aligns with the form postulated in Ref.~\cite{Xing-NC-ST}, provided there is no entropy pumping. The novel contribution of the present work lies in demonstrating that the quantity \( -\beta Q \) precisely corresponds to the change in the Boltzmann entropy of the bath.

Taken together with our previous works, the present theory provides a complete and internally consistent framework for stochastic thermodynamics, valid irrespective of the strength and variation of system–bath interactions and correlations. Remarkably, the resulting theory retains the same structural form as the conventional weak-coupling formalism, with only two key reinterpretations: the system Hamiltonian \( H_X \) must be understood as the Hamiltonian of mean force, and the bath entropy must be defined as the Boltzmann entropy conditional on the system state and the control parameter.

While the microcanonical ensemble is a standard tool in equilibrium statistical mechanics, our use of it in this work is much more refined. Beyond extending the formalism to non-equilibrium settings, we focus on quantities—such as heat, work, and entropy production—that are sub-extensive in the bath size and typically neglected in conventional ensemble theory. Yet in stochastic thermodynamics, it is precisely these sub-extensive contributions that matter, as the extensive parts merely characterize the bulk environment and do not affect dynamical irreversibility. Our framework therefore not only consolidates the thermodynamic theory of small systems but also constitutes a significant advancement in ensemble theory itself, with promising implications for future developments in non-equilibrium statistical mechanics.

The author acknowledges support from NSFC \#12375035.


\begin{thebibliography}{12}

\bibitem{Peliti-book}
L. Peliti and S. Pigolotti, \textit{Stochastic Thermodynamics: An Introduction} (Princeton University Press, 2021).

\bibitem{Jarzynski-review}
C. Jarzynski, ``Equalities and Inequalities: Irreversibility and the Second Law of Thermodynamics at the Nanoscale,'' \textit{Annu. Rev. Condens. Matter Phys.} \textbf{2}, 329–351 (2011).

\bibitem{sekimoto1998} K. Sekimoto, Stochastic energetics of Brownian motion, Phys. Rev. Lett. \textbf{80}, 3424 (1998).

\bibitem{Seifert-review}
U. Seifert, ``Stochastic thermodynamics, fluctuation theorems and molecular machines,'' \textit{Rep. Prog. Phys.} \textbf{75}, 126001 (2012).

\bibitem{seifert2005} U. Seifert, Entropy production along a stochastic trajectory and an integral fluctuation theorem, Phys. Rev. Lett. \textbf{95}, 040602 (2005).

\bibitem{Jarzynski1997}
C. Jarzynski, Nonequilibrium equality for free energy differences, Phys. Rev. Lett. \textbf{78}, 2690 (1997).

\bibitem{Bochkov1977}
G. N. Bochkov and Yu. E. Kuzovlev, General theory of thermal fluctuations in nonlinear systems, Sov. Phys. JETP \textbf{45}, 125 (1977).

\bibitem{Bochkov1981}
G. N. Bochkov and Yu. E. Kuzovlev, Nonlinear fluctuation-dissipation relations and stochastic models in nonequilibrium thermodynamics I, Physica A \textbf{106}, 443 (1981).


\bibitem{Crooks1999}
Crooks, G. E. (1999). Entropy production fluctuation theorem and the nonequilibrium work relation for free energy differences. \textit{Physical Review E}, 60(3), 2721 (1999).

\bibitem{lebowitz1999} J. L. Lebowitz and H. Spohn, A Gallavotti-Cohen-type symmetry in the large deviation functional for stochastic dynamics, J. Stat. Phys. \textbf{95}, 333 (1999).

\bibitem{gallavotti1995} G. Gallavotti and E. G. D. Cohen, Dynamical ensembles in stationary states, J. Stat. Phys. \textbf{80}, 931 (1995).

\bibitem{zwanzig1960} R. Zwanzig, Ensemble method in the theory of irreversibility, J. Chem. Phys. \textbf{33}, 1338 (1960).

\bibitem{mori1965} H. Mori, Transport, collective motion, and Brownian motion, Prog. Theor. Phys. \textbf{33}, 423 (1965).


\bibitem{Zwanzig-book}
R. Zwanzig, \textit{Nonequilibrium Statistical Mechanics} (Oxford University Press, 2001).


\bibitem{Breuer-2005}
H.-P. Breuer and F. Petruccione, \textit{The Theory of Open Quantum Systems} (Oxford University Press, 2002).


\bibitem{Hanggi-2005}
P. H\"anggi and G.-L. Ingold, ``Fundamental aspects of quantum Brownian motion,'' \textit{Chaos} \textbf{15}, 026105 (2005).

\bibitem{Xing-CG}
Ding, Mingnan, and Xiangjun Xing. ``Multi-scale projection operator method and coarse-graining of covariant Fokker-Planck theory." Physical Review Research 5.1 (2023): 013193.

\bibitem{bergmann1955} 
P. G. Bergmann and J. L. Lebowitz, New approach to nonequilibrium processes, Phys. Rev. \textbf{99}, 578 (1955).


\bibitem{Hanggi-Stochastic-review-1982} 
Hänggi, Peter, and Harry Thomas. ``Stochastic processes: Time evolution, symmetries and linear response." Physics Reports 88.4 (1982): 207-319.

\bibitem{Seifert-SC}
Seifert, Udo. ``First and second law of thermodynamics at strong coupling." Physical review letters 116.2 (2016): 020601.

\bibitem{Hanggi-2016-SC}
Talkner, Peter, and Peter Hänggi. "Open system trajectories specify fluctuating work but not heat." Physical Review E 94.2 (2016): 022143.

\bibitem{Jarzynski-2017-SC}
Jarzynski, Christopher. ``Stochastic and macroscopic thermodynamics of strongly coupled systems." Physical Review X 7.1 (2017): 011008.

\bibitem{Esposito-2017-SC}
Strasberg, Philipp, and Massimiliano Esposito. ``Stochastic thermodynamics in the strong coupling regime: An unambiguous approach based on coarse graining." Physical Review E 95.6 (2017): 062101.


\bibitem{Hanggi-2020-comment}
P. Talkner and P. Hanggi. 
``Comment on `Measurability of nonequilibrium thermodynamics in terms of the Hamiltonian of mean force' '' 
Phys. Rev E 21, 066101 (2020)

\bibitem{Hanggi-2020-Colloquium}
Talkner, Peter, and Peter Hanggi. 
``{\it Colloquium:} Statistical Mechanics and Thermodynamics at Strong Coupling:  Quantum and Classical.''  Rev. Mod. Phys. 92, 041002 (2020). 

\bibitem{Xing-SC-2}
Xing, Xiangjun, and Mingnan Ding. ``Thermodynamics and stochastic thermodynamics of strongly coupled systems." Physical Review E 109.3 (2024): 034105.

\bibitem{Xing-SC-1}
Ding, Mingnan, Zhanchun Tu, and Xiangjun Xing. ``Strong coupling thermodynamics and stochastic thermodynamics from the unifying perspective of time-scale separation." Physical Review Research 4.1 (2022): 013015.

\bibitem{Xing-PI}
Ding, Mingnan, and Xiangjun Xing. ``Time-slicing path-integral in curved space." Quantum 6 (2022): 694.

\bibitem{Maes2003}
Maes, Christian, and Karel Netocny. ``Time-reversal and entropy." Journal of statistical physics 110 (2003): 269-310.
\bibitem{Maes2020}
Maes, C. Local Detailed Balance.  \textit{arXiv:2011.09200}.

\bibitem{Leggett-book}
Leggett, Anthony J. Quantum liquids: Bose condensation and Cooper pairing in condensed-matter systems. Oxford university press, 2006.

\bibitem{Xing-NC-ST}
Ding, Mingnan, Fei Liu, and Xiangjun Xing. ``Unified theory of thermodynamics and stochastic thermodynamics for nonlinear Langevin systems driven by non-conservative forces." Physical Review Research 4.4 (2022): 043125.

\bibitem{Xing-ST-cov}
Ding, Mingnan, and Xiangjun Xing. ``Covariant nonequilibrium thermodynamics from Ito-Langevin dynamics." Physical Review Research 4.3 (2022): 033247.

\bibitem{Xing-PI-2022}
Ding, Mingnan, and Xiangjun Xing. ``Time-slicing path-integral in curved space." Quantum 6 (2022): 694.

\bibitem{Talkner2008}
P. Talkner, P. Hänggi, and M. Morillo, 
\emph{Microcanonical quantum fluctuation theorems}, 
Phys. Rev. E \textbf{77}, 051131 (2008).

\bibitem{Talkner2013}
P. Talkner, M. Campisi, and P. Hänggi, 
\emph{Statistics of work and fluctuation theorems for microcanonical initial states}, 
New J. Phys. \textbf{15}, 095001 (2013).

\bibitem{Polkovnikov2011}
A. Polkovnikov, K. Sengupta, A. Silva, and M. Vengalattore, 
\emph{Colloquium: Nonequilibrium dynamics of closed interacting quantum systems}, 
Rev. Mod. Phys. \textbf{83}, 863 (2011).

\bibitem{Goold2016}
J. Goold, M. Huber, A. Riera, L. del Rio, and P. Skrzypczyk, 
\emph{The role of quantum information in thermodynamics---a topical review}, 
J. Phys. A: Math. Theor. \textbf{49}, 143001 (2016).

\bibitem{Vinjanampathy2016}
S. Vinjanampathy and J. Anders, 
\emph{Quantum thermodynamics}, 
Contemp. Phys. \textbf{57}, 545 (2016).

\bibitem{Gogolin2016}
C. Gogolin and J. Eisert, 
\emph{Equilibration, thermalisation, and the emergence of statistical mechanics in closed quantum systems}, 
Rep. Prog. Phys. \textbf{79}, 056001 (2016).

\bibitem{Hanggi-2014-entropy}
Hilbert, Stefan, Peter Hanggi, and Jorn Dunkel. "Thermodynamic laws in isolated systems." Physical Review E 90.6 (2014): 062116.

\bibitem{Hanggi-2016-entropy}
Hanggi, Peter, Stefan Hilbert, and Jorn Dunkel. "Meaning of temperature in different thermostatistical ensembles." Philosophical Transactions of the Royal Society A: Mathematical, Physical and Engineering Sciences 374.2064 (2016): 20150039.

\bibitem{Frenkel-2015-entropy}
Frenkel, Daan, and Patrick B. Warren. "Gibbs, Boltzmann, and negative temperatures." American Journal of Physics 83.2 (2015): 163-170.

\end{thebibliography}
\end{document}